# Optimization of Scheduling in Wireless Ad-Hoc Networks Using Matrix Games

Ebrahim Karami and Savo Glisic, Senior Member IEEE

Centre for Wireless Communications (CWC), University of Oulu,

P.O. Box 4500, FIN-90014, Oulu, Finland

*Abstract—* In this paper, we present a novel application of matrix game theory for optimization of link scheduling in wireless *ad-hoc* networks. Optimum scheduling is achieved by soft coloring of network graphs. Conventional coloring schemes are based on assignment of one color to each region or equivalently each link is member of just one partial topology. These algorithms based on coloring are not optimal when links are not activated with the same rate. Soft coloring, introduced in this paper, solves this problem and provide optimal solution for any requested link usage rate. To define the game model for optimum scheduling, first all possible components of the graph are identified. Components are defined as sets of the wireless links can be activated simultaneously without suffering from mutual interference. Then by switching between components with appropriate frequencies (usage rate) optimum scheduling is achieved. We call this kind of scheduling as soft coloring because any links can be member of more than one partial topology, in different time segments. To simplify this problem, we model relationship between link rates and components selection frequencies by a matrix game which provides a simple and helpful tool to simplify and solve the problem. This proposed game theoretic model is solved by fictitious playing method. Simulation results prove the efficiency of the proposed technique compared to conventional scheduling based on coloring.

## I. INTRODUCTION

Wireless *ad-hoc* networks need a multiple access control scheme to avoid collision due to simultaneous transmissions. The most conventional multiple access scheme in wireless *ad-hoc* networks is time division multiple access (TDMA) [1, 2], although other multiple access schemes like combination of *TDMA* with code division multiple access (CDMA) have been also used. For comparison, TDMA/CDMA is easier to synchronize but provides lower throughputs [3, 4]. In a TDMA wireless *ad-hoc* network, scheduling between links increases throughput. Scheduling in wireless *ad-hoc* networks is similar to channel reusing in cellular networks i.e. if two links do not interfere with each other or at least their mutual interference to signal ratio is less than a predefined margin, they can be activated at the same time slot and consequently not only throughput is increased but also transmission delay is reduced [5, 6]. If two links interfere with each other, they are referred to as adjacent links. Therefore for each graph one adjacency matrix is defined.

In [7, 8] problem of conflict free scheduling has been simplified through splitting network to multiple network realizations or partial network topologies. Each partial topology includes a complete set of links that can be activated simultaneously and consequently to maximize the throughput, the remained problem is just assigning optimum usage rates (portion of time) to each network realization. This approach for scheduling problem has been adopted from coloring problem in mathematics where a graph is divided to some regions and each region is colored with one assigned color. But this approach has two deficiencies. First, coloring does not have unique solution and for any given link adjacency matrix there are usually more than one minimal coloring scheme. On the other hand, when links are supposed to be activated in different usage rates, link coloring does not provide optimum result. To solve both deficiencies we introduce the concept of soft coloring. Soft coloring means more than one color can be assigned to each link in different time slots. The outcome of soft coloring in scheduling problems is dividing topology to topology components where each component is painted by an individual color and components can have non-empty intersection i.e. one link can be member of more than one component in different time slots.

Game theory has found many applications in simplify problems in mathematics and networking [9-11]. In this paper, we use matrix games to model relationship between link usage rates and component rates. Modeling the problem by game theory not only helps us to simplify the problem by using dominancy concept, but also provides fast and reliable convergence.

The rest of paper is organized as follows. In Section II, system model and problem definition is presented. In Section III, problem is formulated by matrix game. Simulation results are presented in Section IV and finally paper is concluded in Section V.

## II. PROBLEM DEFINITION AND SYSTEM MODEL

Assume a multi-source wireless *ad-hoc* network including $N$ nodes. This network is defined as $G(V, E, \zeta, s_1, s_2, \ldots, s_M)$ where $V$ is set of node with $N$ elements, $E$ is set of $L$ virtual wireless links, $\zeta$ is set of $M$ sources and $s_i$ is set of sinks corresponding to the $i$th source where if source is sending unicast data size of its sink set is one. Wireless propagation for this network assumes the following:

1. Omni-directional transmission.
2. Presence of interference due to simultaneous transmission.
3. TDMA as multiple access scheme for different hops without inter time slot interference.

### A. Conflict Free Operation

Assume $S_{ij}$ as the power, in dB required at node $j$, for transmitting node $i$ to reach the receiving node $j$ at distance $d_{ij}$ with $S_{ij} \propto S_t d_{ij}^{-\alpha}$ where $\alpha$ is attenuation factor. Accordingly, by definition of the conflict free scheduling, any node $k \neq i, j$, receiving the signal from node $m$, will be interfered by link $l_{ij}$ if and only if $S_{mk} \leq S_{ik} + \beta$, where $\beta$, in dB, is acceptable interference margin between two links. In other word, link $l_{ij}$ is adjacent to $l_{mk}$ for any $m$ and any $k \neq i, j$ if

$$S_{mk} \leq S_{ik} + \beta. \quad (1)$$

Alternatively, the two links are adjacent if

$$S_{ij} \leq S_{mj} + \beta. \quad (2)$$

Whenever $l_{ij}$ and $l_{mk}$ are physically adjacent i.e. they have a common node or either (1) or (2) hold, they cannot be painted by the same color. Using (1) and (2), link adjacency matrix which is used to design coloring algorithm is defined.

### B. Conventional Scheduling

In general conflict free scheduling, for a given interference margin, is based on conventional graph coloring techniques. A simple and low complexity algorithm for minimal coloring i.e. painting all links with minimum required number of colors is as follows,

Step 1. First link is assigned to the first ($i=1$) partial topology (color), i.e. $T_1=\{l_1\}$.

Step 2. Next unassigned link $l_j$ is chosen.

Step. 3. Using (1), if for any value of $i$, $l_j$ can be simultaneously activated with all members of $T_i$, it is added to $T_i$ as $T_i \leftarrow T_i \cup l_j$.

Step 4. If for all i, where $i \leq i_{max}$, last step cannot be run, $l_j$ is considered as first element of $T_{i_{max}+1}$ and $i_{max} \leftarrow i_{max}+1$.

Step 5. If all links are allocated to $T_i$, we have a complete set of partial topologies and if not, go to Step 2 and run the last two steps for all remained links.

$T_i$s computed by a coloring scheme like this, form a complete set of partial topologies where each link is member of only one of partial topologies.

Following simple example shows inefficiency of coloring when links must be fired with different usage rates.

Assume 3 links $l_1$, $l_2$, $l_3$ with rates $r_1=3$, $r_2=1$, $r_3=2$, where link 1 can be activated with other two simultaneously and two others cannot be activated together. Minimal coloring schemes for these 3 links with their assigned rates are as follows

i) $T_1=\{l_1, l_2\}$ and $T_2=\{l_3\}$ required number of time slots is 5.

ii) $T_1=\{l_1, l_3\}$ and $T_2=\{l_2\}$ required number of time slots is 4.

But obviously none of these coloring schemes are optimum and optimum scheduling for these links is,

iii) $T_1=\{l_1, l_2\}$ during the first time slot and $T_2=\{l_1, l_3\}$ during the next two slots; required number of time slots is 3.

Therefore in optimum solution $T_i$ have non-empty intersection and we call them components of the topology. A component is a set of links that can be activated at the same time. Any single link is also a component and we call them as first generation components and $\tau$th generation of components means set of all components each with $\tau$ members. Components of each generation are parents of one in the next generation. For our 3-links example, we have 5 components as follows,

$T_1=\{l_1\}$, $T_2=\{l_2\}$, $T_3=\{l_3\}$, $T_4=\{l_1, l_2\}$ and $T_5=\{l_1, l_3\}$,

where first 3 components are parents of the last two ones. Consequently in general case to optimize the scheduling appropriate components and their optimal rates must be found.

## III. MATRIX GAME FORMULATION

Given link activation rate vector $r$ with $I$ elements and component set $C$ with $J$ elements, we define following payoff matrix $H$ as follows,

$$h_{ij} = \begin{cases} \frac{1}{r_i}, & if\ l_i \in C_j \\ 0 & otherwise \end{cases} \qquad (3)$$

where $r_i$ is activation rate for $i$th link and $C_j$ is $j$th component. Assume $H$ as payoff matrix of the min-max zero sum game between two players where their strategy sets are links and components sets respectively. In the sequel we prove that the mixed strategy vector of the second player gives optimum component rate.

*Theorem-* mixed equilibrium of the zero sum game defined by payoff matrix (3) gives optimum scheduling for rate vector $r$.

Proof: Assume $x$ and $y$ as mixed strategies of the players at equilibrium. Since $y$ is normalized to 1 activation of components in average needs one time slot. $\tilde{r}_i$ as activation rate of $i$th link supported by $y$ components is computed as follows,

$$\frac{\tilde{r}_i}{r_i} = {}_i[H]y \qquad (4)$$

where ${}_i[H]$ is $i$th row of the $H$. Therefore by component rate vector $y$, the link with minimum supported rate which is bottleneck of the game is,

$$i_{minimum} = \arg\min{}_i \left({}_i[H]y\right) \qquad (5)$$

Consequently optimum component rate vector must maximize $\min{}_i \left({}_i[H]y\right)$ a

$$y = \arg\max{}_y \min{}_i \left({}_i[H]y\right) \qquad (6)$$

And theoretically (6) is equivalent to

$$y = \arg\max{}_y \min{}_x \left(x^T H y\right), \qquad (7)$$

where (7) defines equilibrium for game defined by payoff matrix $H$ and value $\max{}_y \min{}_x \left(x^T H y\right)$.

Formulation of the problem as a game gives us the chance to simplify the model using especial properties of the games, like dominancy theory. For instance in general case while solving this game, because of dominancy, components tagged as parent can be ignored because apparently any parent is dominated by its child. Therefore we just need to consider last generation generated born from any link. On the other hand, calculation of mix strategy for linear games is easy and can be performed by fictitious playing method (FP). FP property of the games is a feature of stationary games where players can update their belief on other player's strategies based on history of their decisions. This technique which is used to solve matrix games, was proposed by Brown [13] and its convergence for different conditions was proved in [14-19]. FP algorithm for a min-max zero-sum game has following steps,

Step 1. Initialization of $\hat{x}$ as,

$$\hat{x}_0 = {}_i[H]^T, \qquad (8)$$

where ${}_i[H]$ is and arbitrary row of $H$ ($\hat{x}$ actually will be $Hy$). Then set iteration number $k=1$.

Step 2. Finding best strategy for the first player at $k$th iteration as,

$$i_k = \arg\min{}_m \hat{x}_{k-1,m} \qquad (9)$$

where $\hat{x}_{k,m}$ is $m$th element of the $\hat{x}_k$.

Step 3. Updating $\hat{y}_k$ as, $\hat{y}_k = \hat{y}_{k-1} + [H]_{i_k}$.

Step 4. Finding best strategy for the second player at $k$th iteration as,

$$j_k = \arg\max{}_n \hat{y}_{k,n} \qquad (10)$$

where $\hat{y}_{k,n}$ is $n$th element of the $\hat{y}_k$.

Step 5. If We algorithm has not converged to equilibrium set $k \leftarrow k+1$ then update $\hat{x}_k$ as,

$$\hat{x}_k = \hat{x}_{k-1} + {}_{j_k}[H], \qquad (11)$$

and then return to Step 2. Algorithm is $\delta$ converged if,

$$\hat{y}_{k,j_k} - \hat{x}_{k,i_k} \leq \delta, \qquad (12)$$

When the algorithm approaches equilibrium mixed strategies for both players are calculated by averaging over best strategies calculated at each iteration.

## IV. SIMULATION RESULTS

The proposed algorithm is simulated over randomly generated graphs with nodes uniformly distributed over unit square. Sources and sinks are also selected randomly and for each source-sink pair shortest path (physically shortest part that requires minimum power) is calculated using Dijkstra algorithm [20]. Attenuation factor $\alpha$ is assumed to be 4 and number of packets to deliver from each source to its corresponding sink is randomly chosen by Poisson distribution.

The proposed optimal algorithm is compared to the conventional scheduling based on network graph coloring and no scheduling case. The average number of required time slots to carry each packet from source to its corresponding sink is considered as performance criterion and results are averaged over 1000 independent runs. Results are summarized in Figs. 1-4.

Fig. 1, presents average number of required time slots versus interference margin ($\beta$) for different numbers of source-sink sessions when $N$ =10 nodes. We can see from the figure that the proposed scheduling outperformance conventional scheme. In the case where we have 10 parallel source-sink sessions, optimum scheduling offers more scheduling gain compared to 5 source-sink sessions because when higher number of paths must be scheduled, more paths is participating in the scheduling process. Therefore every link has more partners for simultaneous transmission. In other word, number of last generation of components is higher and consequently scheduling gain is also higher.

Fig. 2, presents the same results for N=20. In this case, when we have 10 parallel source-sink pairs, both optimum and conventional scheduling offer more scheduling gain compared to the 5 parallel pairs but gain of the optimum scheduling is still much higher.

Figs. 3 and 4 present average number of required time slots versus number of parallel source-sink pairs. We can see that the scheduling gain advantage of optimum scheduling compared to the conventional scheduling increases with the number of source-sink pairs. For instance, in the case of 10 source-sink pairs, optimum scheduling needs between 11 and 25 percent less time slots in average.

## V. CONCLUSION

In this paper a novel technique based on matrix games is proposed to solve optimal scheduling problem is wireless *ad-hoc* network. In the proposed method, first wireless topology is divided to components which actually are overlapped partial topologies. Then by assigning appropriate usage rates to the selected components optimum scheduling is achieved. The optimal rates for graph components are obtained by calculating mix equilibrium of a matrix game between links and components. Using matrix games for this kind of problem helps us to simplify the problem through special features of the games, and using fictitious playing method which works iteratively, provides quick convergence. The proposed game model has been simulated for multiple-unicast sessions and the results are averaged over 1000 independent runs. Simulation results prove the efficiency of the proposed method compared with the conventional scheduling.

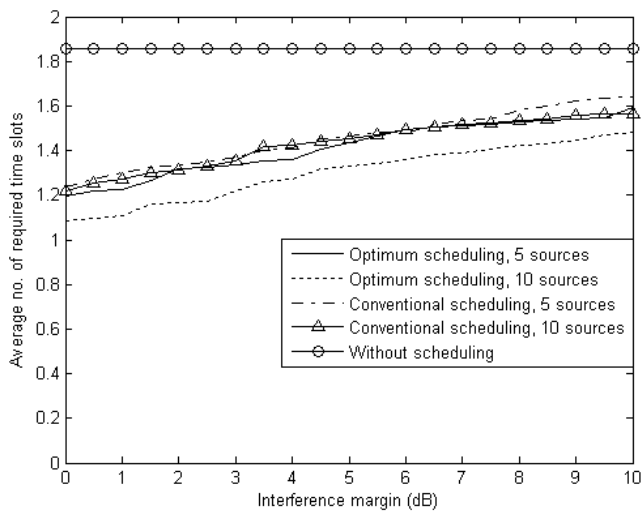

Fig. 1. Average no. of required time slots w.r.t. interference margin for N=10.

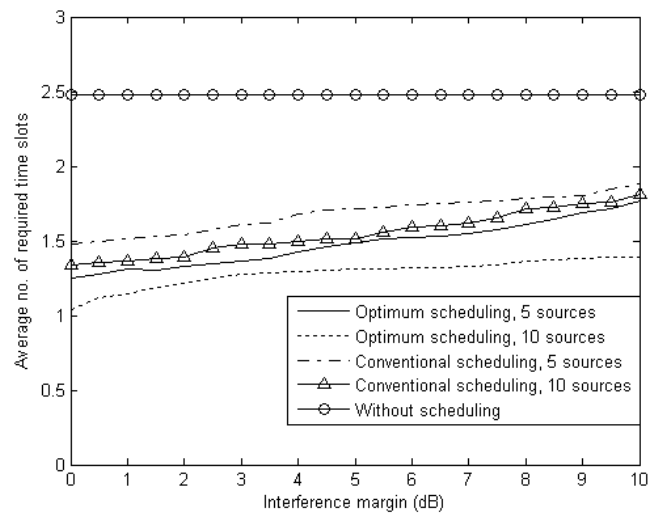

Fig. 2. Average no. of required time slots w.r.t. interference margin for N=20.

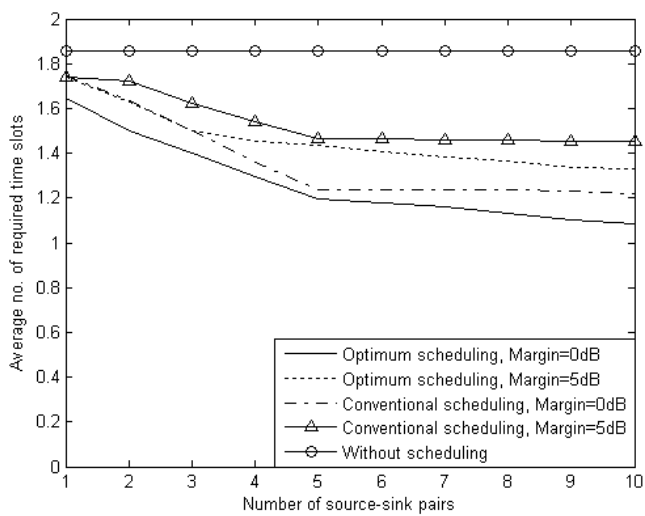

Fig. 3. Average no. of required time slots number f source-sink pairs for N=10.

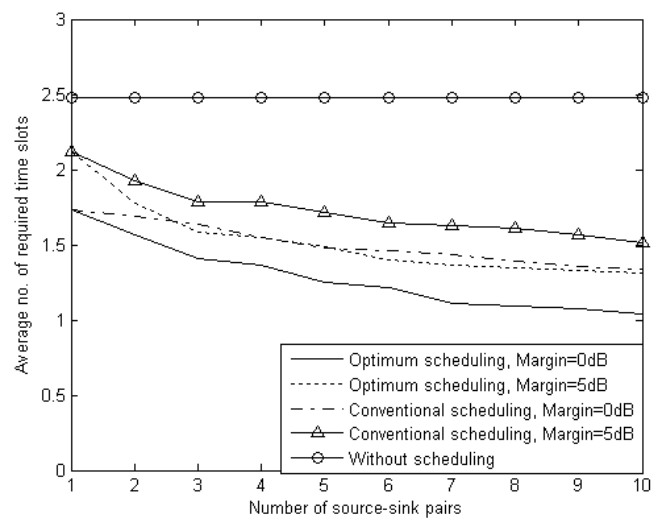

Fig. 4. Average no. of required time slots number f source-sink pairs for N=20.